%% file: main.tex
\documentclass[sigconf]{acmart}

\AtBeginDocument{%
  \providecommand\BibTeX{{%
    \normalfont B\kern-0.5em{\scshape i\kern-0.25em b}\kern-0.8em\TeX}}}

\copyrightyear{2024}
\acmYear{2024}
\setcopyright{rightsretained}
\acmConference[CF '24]{21th ACM International Conference on Computing Frontiers}{May 7-9, 2024}{Ischia, Italy}
\acmBooktitle{21th ACM International Conference on Computing Frontiers (CF '24), May 7--9, 2023, Bologna, Italy}

\usepackage[per-mode=symbol,retain-explicit-plus,separate-uncertainty=true]{siunitx}
\usepackage{url}
\usepackage{graphicx, nicefrac}
\usepackage{lipsum}
\usepackage{hyperref}
\usepackage{dblfloatfix}    
\usepackage{enumitem}
\usepackage{caption}
\usepackage{titlesec}
\usepackage{natbib}

\titlespacing{\subsubsection}{0.4pt}{\parskip}{-\parskip}
\titlespacing{\subsection}{0.1pt}{\parskip}{-\parskip}

\newcommand{\oecgra}{OpenEdgeCGRA}
\newcommand{\imic}{Im2col\mbox{-}IP}
\newcommand{\imoc}{Im2col\mbox{-}OP}
\newcommand{\convoc}{Conv\mbox{-}OP}
\newcommand{\heepsilon}{$H\mathcal{E}\mathcal{E}Psilon$}
\newcommand{\Heepsilon}{$\mathbf{H\mathcal{E}\mathcal{E}Psilon}$}
\newcommand{\energyimprov}{$3.4\times$}
\newcommand{\latencyimprov}{$9.9\times$}

\newcommand{\xheep}{X\mbox{-}HEEP}

\newcommand{\Secref}[1]{Section~\ref{#1}} 

\usepackage[acronym, toc,numberedsection = nolabel, section = part]{glossaries}
\makenoidxglossaries
\loadglsentries{Refs/gls}
\glsdisablehyper
\renewcommand*{\a}[1]{\gls{#1}}
\newcommand*{\as}[1]{\glspl{#1}}

\begin{document}

\title{Performance evaluation of acceleration of convolutional layers on \oecgra{}}



\author{Nicolò Carpentieri$^1$, Juan Sapriza$^2$, Davide Schiavone$^2$, Daniele Jahier Pagliari$^1$, David Atienza$^2$, Maurizio Martina$^1$, Alessio Burrello$^1$}
\affiliation{ \institution{$^1$ Politecnico di Torino, Torino, Italy, $^2$EPFL, Lausanne, Switzerland} \country{} }

\settopmatter{printacmref=false}






\email{juan.sapriza@epfl.ch}

\renewcommand{\shortauthors}{Carpentieri, et al.}


\begin{abstract}

    Recently, efficiently deploying deep learning solutions on the edge has received increasing attention. New platforms are emerging to support the increasing demand for flexibility and high performance.
    In this work, we explore the efficient mapping of convolutional layers on an open-hardware, low-power \a{cgra}, namely \oecgra{}.
    We explore both direct implementations of convolution and solutions that transform it into a matrix multiplication through an Im2col transformation, and experiment with various tensor parallelism axes.
    We show that for this hardware target, direct convolution, coupled with weight parallelism reaches the best latency and energy efficiency, outperforming a CPU implementation by \energyimprov{} and \latencyimprov{} in terms of energy and latency, respectively.
    %
    %

\end{abstract}

\maketitle

\input{Sections/01_intro} 
\input{Sections/02_methods}

\input{Sections/03_Results}

\input{Sections/04_Conclusion}


\bibliographystyle{unsrt}
\bibliography{main}

\end{document}

%% file: Refs/gls.tex
\newacronym{alap}{ALAP}{As-Late-As-Possible}
\newacronym{asap}{ASAP}{As-Soon-As-Possible}
\newacronym{alu}{ALU}{Arithmetic-Logic Unit}
\newacronym{asic}{ASIC}{Application Specific Integrated Circuit}
\newacronym{cgra}{CGRA}{Coarse-Grain Reconfigurable Array}
\newacronym{cnf}{CNF}{Conjunctive Normal Form}
\newacronym{cnn}{CNN}{Convolutional Neural Network}
\newacronym{cil}{CIL}{Compute-Intensive Loop}
\newacronym{dfg}{DFG}{Data Flow Graph}
\newacronym{dma}{DMA}{Direct Memory Access}
\newacronym{fpga}{FPGA}{Field Programmable Gate Array}
\newacronym{ims}{IMS}{Iterative Modulo Scheduling}
\newacronym{isa}{ISA}{Instruction Set Architecture}
\newacronym{ilp}{ILP}{Integer Linear Programming}
\newacronym{ir}{IR}{Intermediate Representation} 
\newacronym{ii}{II}{Iteration Interval}
\newacronym{im2col}{Im2col}{Image-to-Column}
\newacronym{ip}{IP}{Input-Channel Parallelism}
\newacronym{kms}{KMS}{Kernel Mobility Schedule}
\newacronym{mii}{mII}{minimum II}
\newacronym{ms}{MS}{Mobility Schedule}
\newacronym{mac}{MAC}{Multiply-and-Accumulate}
\newacronym{op}{OP}{Output-Channel Parallelism}
\newacronym{os}{OS}{Output Stationary}
\newacronym{pc}{PC}{Program Counter}
\newacronym{pe}{PE}{Processing Element}
\newacronym{ra}{RA}{Register Allocation}
\newacronym{rf}{RF}{Register File}
\newacronym{rtl}{RTL}{Register Transfer Level}
\newacronym{sat}{SAT}{satisfiability}
\newacronym{soa}{SoA}{State of the Art}
\newacronym{u}{U}{Utilization}
\newacronym{ws}{WS}{Weight Stationary}
\newacronym{wp}{WP}{Weight Parallelism}

%% file: Sections/01_intro.tex
\section{Introduction}

\begin{figure}[t]
    \centering
    \includegraphics[width=1\linewidth]{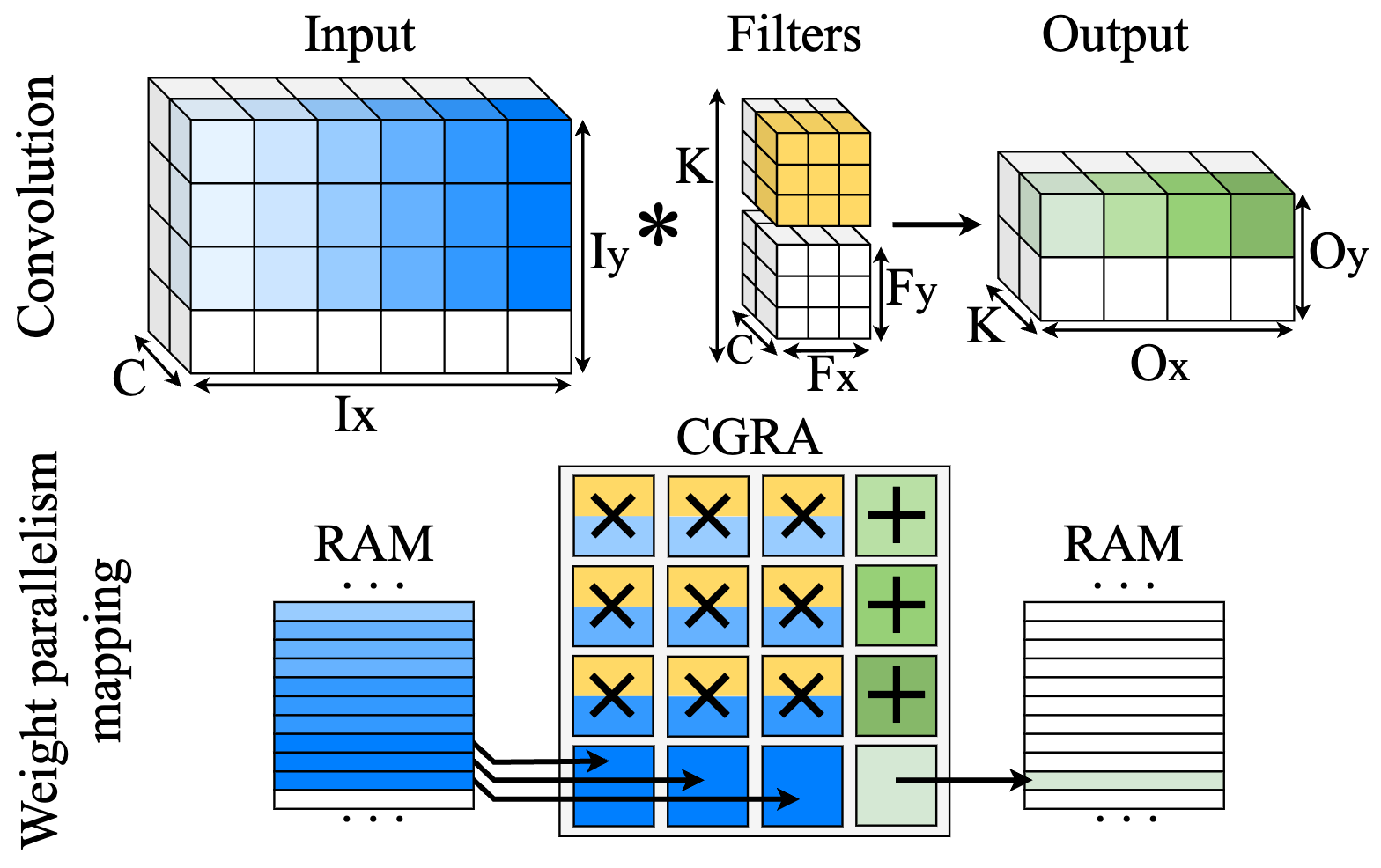}
    \caption{(Top) 2D convolution scheme. (Bottom) Direct convolution with weight parallelism. Nine PEs perform dot products between constant weights and sequentially loaded inputs. The other PEs load new inputs or sum partial outputs.}
    \label{fig:intro}
\end{figure}

\glsresetall{}
The shifting paradigm from cloud to edge computing led to executing deep learning models into low-power and memory-tight devices.
Among the plethora of models, \as{cnn} have emerged as one of the most powerful tools for various tasks, including image recognition, and natural language and signal processing. However, their efficient execution on resource-constrained edge devices remains a significant challenge given the high memory footprint and the huge number of operations.
To cope with this challenge, various architectures have been proposed. ASICs offer the highest performance and energy efficiency for specific applications due to their custom-tailored hardware design~\citep{JouppiYPPABBBBB17}. FPGAs, on the other hand, provide increased flexibility through their fine-grained reconfigurable architecture, making them more versatile than ASICs but generally less energy efficient~\citep{Li2017AGF}.
In this work, we explore the use of \as{cgra} as potential candidates to cover a different part of the design space of edge computing systems (i.e., considering trade-offs between performance, energy efficiency, area and versatility) to execute \a{cnn} applications~\citep{lee2021specializing}.
\as{cgra} present programmable hardware that can be customized to specific tasks, making them well-suited for dynamic edge computing environments. Thus, they offer an attractive meet-in-the-middle solution between performance, efficiency, and task-versatility. Nevertheless, the efficient mapping of convolutional layers onto \as{cgra} is essential to exploit their potential benefits. \autoref{fig:intro} shows an example of the mapping of a convolution on our target CGRA. 

Although several works have focused on mapping techniques for high performance \as{cgra}~\cite{wijerathne2021himap} or specializing \as{cgra} for deep learning~\cite{lee2021specializing,heidorn2020design}, low-power ($mW$-power), general-purpose \as{cgra} have not been explored equally for tinyML~\cite{podobas2020survey}. To fill this gap, this work addresses the problem of mapping convolutions, focusing on the \oecgra{} architecture for edge computing applications~\cite{alvarez2023open}. 
Our goal is to outline efficient practices that allow to leverage this accelerator, minimizing the impact of the overheads it imposes.
For this reason, we investigate various state-of-the-art computational and memory management strategies, aiming to uncover the most efficient mapping technique that balances performance and resource constraints.
Specifically, we present a two-fold contribution: (i) we explore different implementation paradigms for convolution and different tensor parallelism axes; (ii) we benchmark the results of the different implementations, measuring energy, latency, performance, and memory usage, and provide insights on the best mapping technique for low-power CGRA.  
This analysis highlights the predominance of direct convolution, coupled with weight parallelism, which reaches up to \energyimprov{} and \latencyimprov{} in terms of energy and latency, respectively, compared to a plain CPU implementation, achieving an overall average performance of \mbox{$0.6 \nicefrac{MAC}{cycle}$}.

%% file: Sections/02_methods.tex
\section{Material and Methods}\label{sec:methodology}
In this section, we first introduce our target CGRA. Then, we describe the different convolutional kernels mapping techniques explored to exploit its hardware resources maximally.
\begin{figure}[t]
    \centering
    \includegraphics[width=0.95\linewidth]{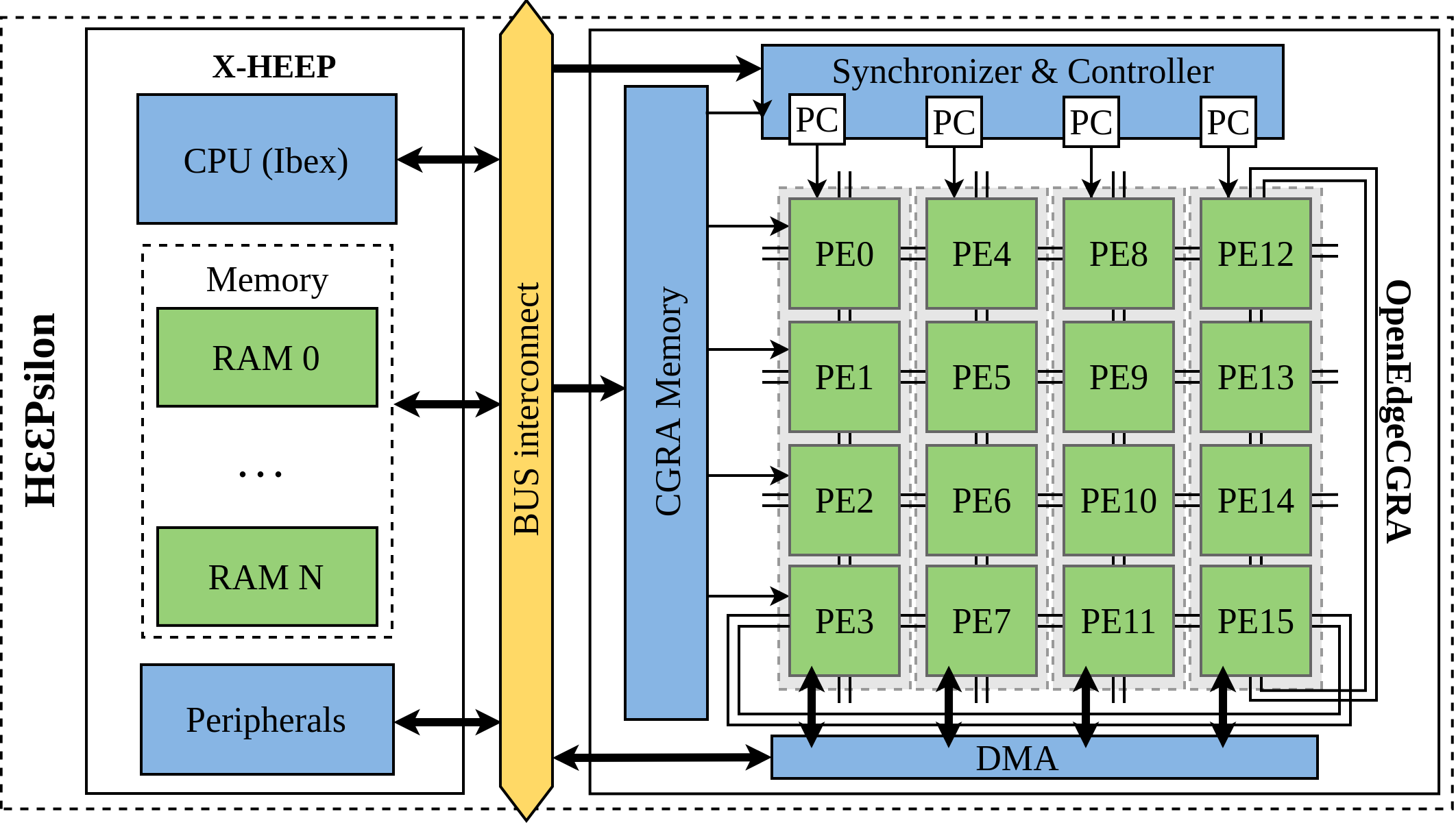}
    \caption{Architecture of the  \Heepsilon{} platform used as a test bench for this analysis, where the \oecgra{} is instantiated along with \xheep{}.}
    \label{fig:heepsilon}
    \vspace{-0.2cm}
\end{figure}

\subsection{Hardware Platform}
In this work we target the \oecgra{} architecture: a low-power, general purpose, scalable, instruction-based \a{cgra}. Although designed to execute health applications in low area and power footprint, thus not ideal for running regular high intensity kernels such as CNNs, this \a{cgra} has been chosen for being open source, validated in silicon, and already integrated into a RISC-V microcontroller~(X\mbox{-}HEEP) in the \heepsilon{} platform.
All our work is open-source\footnote{Available on \url{github.com/esl-epfl/HEEPsilon/tree/convolution_exploration}}.

\autoref{fig:heepsilon} illustrates this platform, which includes a CPU and memory allowing for real application evaluation~\cite{machetti2024x}. The CPU configures the \a{cgra} and loads instructions into the \a{cgra}'s memory before launching its first execution. The instruction set of the \a{cgra} includes 32-bit integer arithmetic and logical operations and supports both conditional and unconditional jumps, allowing the implementation of deep learning kernels. However, a \a{mac} instruction which could lead to higher performance is not supported. In this work, we used an instance of the \oecgra{} with a $4\times4$ matrix of \as{pe}. Each \as{pe} is composed of one \a{alu}, two multiplexed inputs, one output register, a four-element \a{rf}, and a 32-word private program memory. 
The \as{pe} are connected to their neighbors through a torus interconnect, allowing for reuse of neighbor data. 
Although columns in \oecgra{} have their own independent \a{pc}, we always used the four columns as part of a single application, which means that the latency of execution of a single \a{cgra}-instruction is determined by the latency of the slowest operation among the 16 \as{pe}. 
Each column of the \oecgra{} has a port connected to a \a{dma} block, allowing them to access the memory subsystem.

\subsection{Convolution mapping strategies}\label{sec:mappings}
Our work explores different optimization directions to map a convolutional kernel onto the \oecgra{}. 
In particular, we first explore the implementation paradigm, i.e., the data layout coupled with its access order. Then, we investigate the computation's parallelization over the \oecgra{}'s PEs.
For all our experiments, we always consider convolutions with groups = 1, and a filter of dimension $F_X \times F_Y = 3 \times 3$ \cite{bengio2017deep}.

\hfill\break
\textbf{Convolution implementation: \a{im2col} vs Direct Access} 

We consider two different implementations, i.e., the direct convolution and the so-called \a{im2col} transformation. 
The direct convolution does not manipulate the input data. It directly fetches data from memory, leading to non-sequential load operations of the input image and, therefore, higher overhead in data addressing.
To minimize this overhead a Channel-Height-Width (CHW) data layout is typically used \citep{abs-1801-06601}.
On the other hand, the \a{im2col} transformation is the most adopted implementation in CPU and GPU kernel libraries, such as PULP\mbox{-}NN~\cite{garofalo2020pulp}, Mxnet~\citep{ChenLLLWWXXZZ15}, or Tensorflow~\citep{3352022}. It transforms multi-channel 2D Convolutions into a vector-matrix product by turning each input activations' patch (originally a 3D tensor) into a 1D vector of dimension input channels~(C) $\times$ filter rows~($F_X$) $\times$ filter columns~($F_Y$), which is multiplied with the 2D weights matrix, of dimension C$\times F_X\times F_Y\times$ output channels (K); note that this transformation simplifies the memory accesses, which become sequential.
On the other hand, it requires more memory to store the buffer of reordered inputs and additional instructions to create this buffer, which could be non-negligible.
We argue that \a{im2col} transformation can leverage the loads with automatic index increment and parallel DMA ports of our target architecture.
In \citep{abs-1801-06601}, the authors show that the Height-Width-Channel (HWC) data layout is the most advantageous for the creation of the \a{im2col} reorder buffer. Hence, we select it for our implementation.

\textbf{Parallelization Axis} 

The main computation in a \a{cnn} involves six nested loops, which can be swapped and parallelized without changing the final result~\cite{garofalo2020pulp}. These loops correspond to i)~output channels~(K), ii)~input channels~(C), iii)~output rows~($O_X$), iv)~output columns~($O_Y$), v)~filter rows~($F_X$), and vi) filter columns ($F_Y$). 
We explore the parallelization of the C, K, or $F_{X/Y}$ loops. Note that we do not explore parallelization of $O_{X/Y}$ loops given that it would enable reuse of neither the weights nor the inputs.

\textit{\a{wp}}: This method leverages the parallelization of the filter loops, $F_X$ and $F_Y$. In this setup, each weight element of a single input and output channel is assigned to a different PE. For a 3$\times$3 filter, this means that nine weight elements are distributed across nine PEs. Once these weights are retrieved from memory, the system performs multiple \a{mac} operations by updating the inputs for each PE. The partial outputs generated by the PEs then move through the spatial array of the \a{cgra}.
This procedure is illustrated in ~\autoref{fig:intro}. In addition to the nine PEs engaged in computations, the final row (comprising three PEs) is tasked with updating the address to load the new input triplet (3$\times$1), while the other 3$\times$2 inputs can be efficiently reused by shifting them from the first two rows of PEs when computing the next output pixel on the same output image row. During this stage, the last column of PEs aggregates the nine computed partial sums and, if necessary, adds them to a previous partial sum when processing input channels $c_i > 0$. The last PE is designated for storing the accumulated partial sum in memory. This cycle is repeated for the entire input spatial position before a new set of weights is loaded to process the next input channel. The outputs are sequentially generated starting from $O_X$, $O_Y$, and, finally, $K$. Importantly, this mapping scheme, which benefits from a CHW input layout, would not benefit from using the Im2col transformation.

\textit{\a{ip}}: This method involves performing \a{mac} operations relative to various input channels in parallel. It utilizes the \a{im2col} technique to enable sequential access to the input and filter data. Note that using direct convolution for this parallelism strategy would be suboptimal given that the latency to access the data from each single PE would strongly increase given their storage position. In this mapping strategy, for each iteration of the most external loops (K, $O_X$, $O_Y$), every \a{pe} handles a distinct set of input channels (C/16 per PE) for the same output channel and spatial position. In the end, the partial sums computed by each \a{pe} are aggregated, and the next element is computed. 
    
\textit{\a{op}}: This mapping aims to produce results simultaneously for different output channels. Its rationale is to minimize the latency for reading and writing partial sums by keeping them in the \a{rf} of each \a{pe} ~\citep{sze-dnn}. A different output channel is then assigned to each different PE, which stores a different set of weights, and the same input elements are broadcast to all \as{pe}, to produce 16 output channels at the same spatial location in parallel. For this implementation, both the \a{im2col} approach and the direct convolution implementations are considered.

\begin{figure}[!b]
    \centering
    \includegraphics[width=\linewidth]{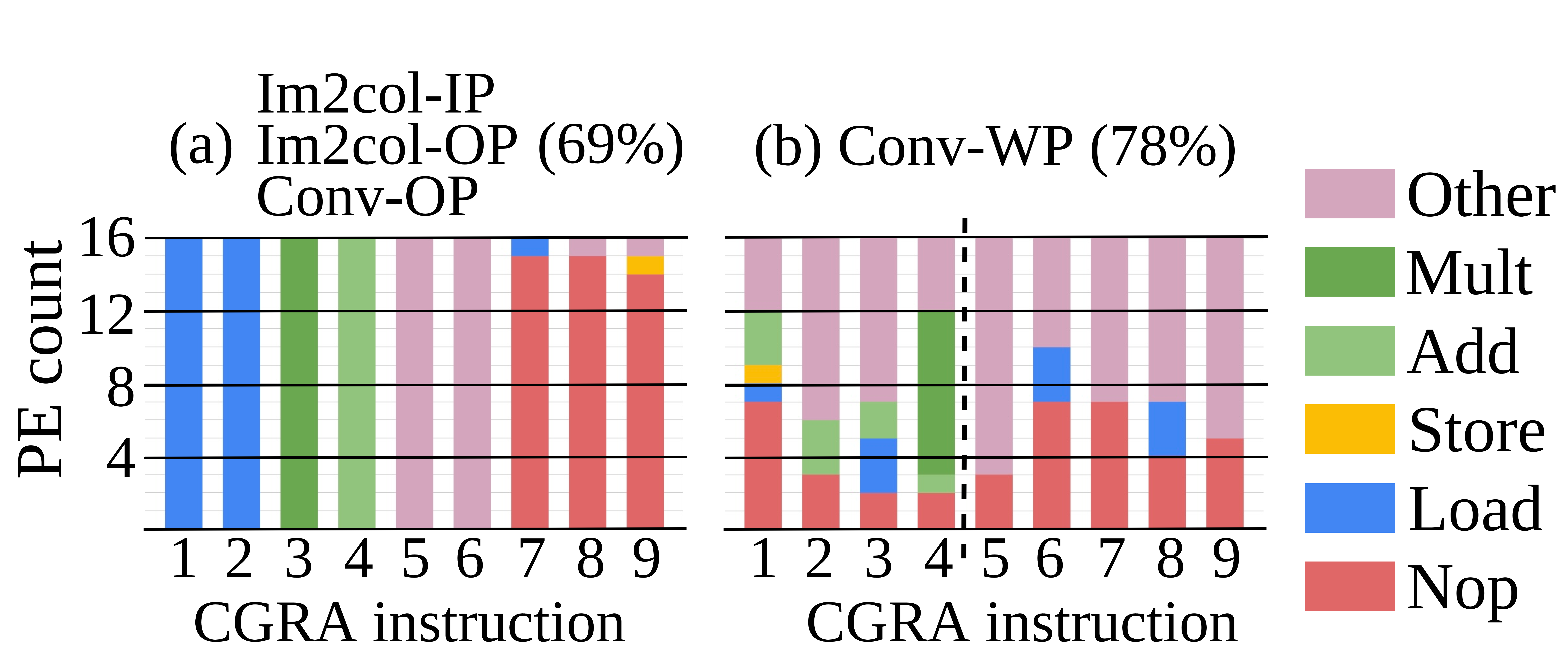}
    \caption{Operation distribution of different convolution mapping strategies. \textit{Other} includes index updates, branch operations, and index manipulation.}
    \label{fig:kernels}
\end{figure}

\autoref{fig:kernels} shows the distribution of the operations in the innermost loop of each mapping strategy, over the CGRA's PEs.
Note that while WP differs significantly, the other three mappings are identical:
in the first two instructions, 16 inputs and weights are loaded (corresponding to 16 input or output channels). Next, the \texttt{mul} and \texttt{sum} operations are executed by all PEs. Then, in the last 5 instructions, the input and weight addresses and the iteration counter are updated, followed by the loop's branch instruction.
%
%
Most PEs execute a \texttt{nop} during the last three instructions because only one to two PEs are in charge of updating the iteration counter and branching for the whole CGRA.
Because of this bottleneck, the innermost loop reaches an overall PE utilization of 69\%.
For all three mapping strategies, this loop is repeated \mbox{$F_X\times F_Y \times O_X \times O_Y \times C \times K / 16$ times}.

Conversely, the WP mapping is composed of a main internal loop and a border internal loop.
The main loop is composed of only 4 instructions that allow the execution of the nine multiplications, the sum reduction, the load of a new input triplet, and the final store. However, as mentioned above, once a new output row has to be processed, also the other 6 inputs (2$\times$3) should change, necessitating 5 additional instructions (border loop) to load the additional data and update indexes, as shown in the graph.
In this case, the main loop is executed $O_X \times O_Y \times C \times K$ times with an utilization of 78\%, while the border one is executed only once per row, \mbox{i.e., $O_Y \times C \times K$ times}.

\subsection{Evaluation Metrics}\label{sec:evaluations}
We compare our convolution mapping strategies in terms of the following metrics to provide a complete overview of the utilization and efficiency of the platform:
%

\textbf{Latency}: the time required to perform a complete convolution, comprising both the \a{im2col} creation (if needed) and the kernel execution. The time required to load the instructions before the first iteration is neglected.
    
\textbf{Energy}: we consider the power consumption of a complete minimal system, including CGRA, CPU and memory subsystems. This allows for a fair comparison between different strategies but should not be used as a benchmark to compare the platform's efficiency, as this was not optimized. In the \a{im2col} case, the MCU performs data reordering during the \a{cgra} execution. At all other times the MCU enters a busy loop waiting for the \a{cgra} interrupt.
    
\textbf{Memory usage}: to assess the scalability of each strategy, we characterize its memory footprint as the space required to store the input and output samples and the weight filters.
%

\textbf{MAC/cycle:} to compare the obtained execution speed with other state-of-the-art implementations,  we compute the performance in terms of \a{mac} operations per clock cycle ($\nicefrac{MAC}{cycle}$).  

Behavioral simulation of each approach was performed using the \oecgra{} simulator\footnote{Available on \url{github.com/esl-epfl/ESL-CGRA-simulator}}. Latency measurements were obtained from the FPGA implementation of \heepsilon{} and validated against pre-synthesis simulation. The average power was obtained from post-synthesis simulation on a TSMC \SI{65}{\nano\meter} technology process, where the \a{cgra} required an area of $\sim\SI{0.4}{\milli\meter\squared}$.

%% file: Sections/03_Results.tex
\section{Experimental results}\label{sec:results}
In this section, we present results concerning the latency and energy efficiency of the kernels described in Section 2, with a final analysis of their robustness to hyper-parameters variations. All kernels use 32-bit integer data.
\begin{figure}[!t]
    \centering
    \includegraphics[width=\linewidth]{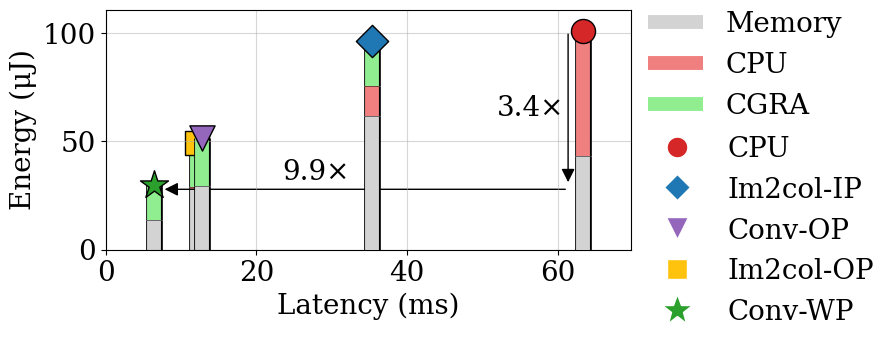}
    \caption{Energy vs. Latency comparison.}
    \label{fig:energy_vs_latency}
\end{figure}

\begin{figure*}[ht]
    \centering
    \includegraphics[width=0.95\textwidth]{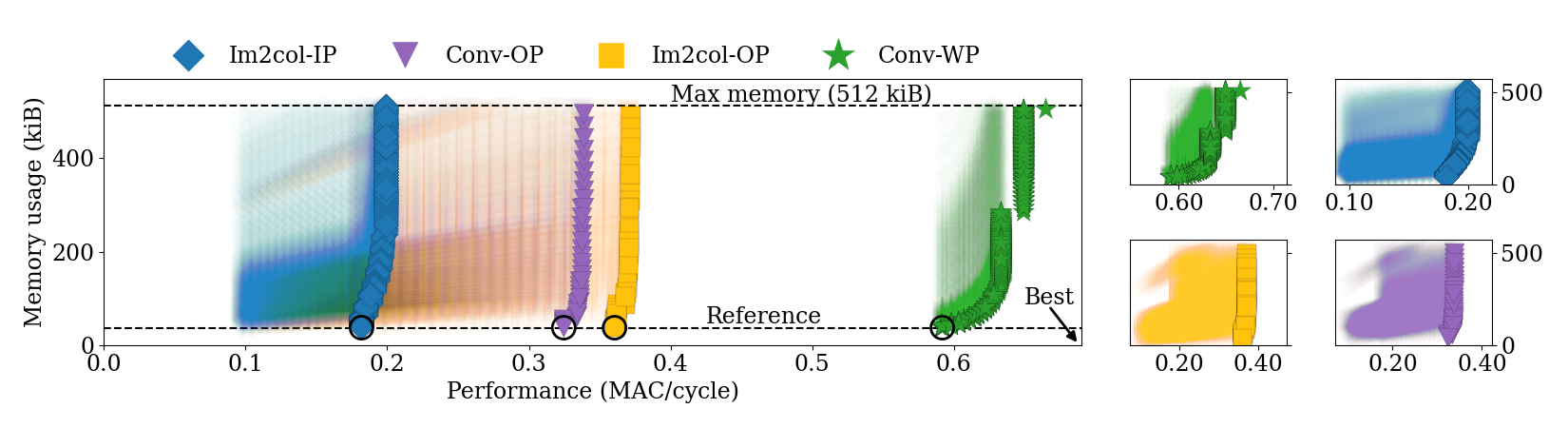}
    \caption{Impact on memory and performance of different hyperparameters. Pareto-optimal results are highlighted with a greater color intensity. The experiments of \Secref{sec:runtime} are highlighted by black circles.}
    \label{fig:mem_vs_mac_ext}
\end{figure*}

\subsection{Energy and latency evaluation}\label{sec:runtime}
We run a baseline convolution with C=K=$O_X$=$O_Y$=16, and a $3\times3$ filter. For each mapping method, we measure execution latency and energy consumption of the three main blocks involved: CGRA, CPU, and memory. In \autoref{fig:energy_vs_latency} results are compared against a CPU-only implementation. The \a{wp} approach reaches energy and latency improvements of \energyimprov{} and \latencyimprov{}, respectively, at an average power of \SI{2.5}{\milli\watt}, the highest among the \a{cgra}-approaches. Its main advantage over the other strategies lies in the high data reuse rate of the weight stationary strategy, which brings two-fold benefits. First, the reduced number of memory accesses and their distribution over time avoids collisions between PEs, hence decreasing the latency of the whole computation. Secondly, the lower number of memory accesses also reduces the dynamic energy consumed by the memory subsystem. \autoref{fig:energy_vs_latency} highlights that the latter is the largest energy-wise discriminative factor between methods -which would have been overlooked by an isolated analysis of the CGRA performance-. 
For example, the higher energy consumption of the \imoc{} approach is not associated with the CPU involvement (which includes the implementation of \a{im2col}) but with the more frequent load instructions. In this case, the CPU contribution is negligible and allows marginal improvements in both energy and latency with respect to the \convoc{} approach. In contrast, the \imic{} method requires frequent computation of the \a{im2col}, leading to a higher CPU activity and doubling memory consumption. This situation also increases latency due to the overhead of launching each iteration. In this method, every call of the \a{im2col} function creates one output position at a time and, additionally, each \a{im2col} input organization has to be repeated for every output channel. Instead, \imoc{} operates in parallel across output channels, allowing it to generate 16 output positions simultaneously with just one \a{im2col} setup.

\subsection{Robustness evaluation}
We evaluate the performance deviation from the \textit{baseline} case explored in the previous section by swiping the layer hyperparameters. We vary $O_X$ and $O_Y$ in $[16, 64]$, C and K in $[16, 144]$, increasing by 1 the dimension of each parameter until 32, and then in steps of 16 given the similar scalability. We limit our search to the maximum memory available in the system (\SI{512}{kiB} from \heepsilon{}'s RAM banks). 
The results, illustrated in \autoref{fig:mem_vs_mac_ext}, show that \a{wp} has the greatest robustness to hyperparameter changes, with increasing layer dimensions always leading to improved performance.  \a{wp} remains the best approach for any hyperparameter combination, reaching up to $0.665 \nicefrac{MAC}{cycle}$ with C =16, K =16, and $O_X = O_Y$ = 64. It is noteworthy that increasing $O_X$ and $O_Y$ translates into an improvement in performance for the \a{wp} case thanks to two different contributions: first, the larger the input size, the higher the reuse of the loaded weights; second, a larger feature map reduces the occurrence of row changes while swiping the input activations, thus, the associated overhead of border loop (cf. Section~\ref{sec:mappings}).

On the other hand, all the other approaches see a drop in performance every time their parallelization dimensions are not a multiple of the number of PEs (i.e., 16), reaching their lowest performance ($\sim0.1 \nicefrac{MAC}{cycle}$) when the parallelization dimension is equal to 17 due to the strong imbalance in the workload distribution. In this case, the \imoc{} results the least robust with a performance reduction of 3.62$\times$ when compared to its best case.

%% file: Sections/04_Conclusion.tex
\section{Conclusions}\label{sec:conclusions}

This work has thoroughly analyzed the impact of different convolution mapping techniques, typical in CNN applications, on the \oecgra{}. Latency, energy, memory usage, and performance were evaluated to conclude that the \a{wp} approach is the best performing one, with a peak of $0.665 \nicefrac{MAC}{cycle}$. Energy and latency improvements compared to the CPU-only implementation reach \energyimprov{} and \latencyimprov{}, respectively. 
While specialized architectures in the state of the art reach $23.3\times$ higher performance~\cite{garofalo2020pulp}, this work shows how to leverage the existing trade-offs between performance, smaller area (\SI{0.4}{\milli\meter\squared}) and low-power ($<\SI{2.5}{\milli\watt}$) on \as{cgra}.
Thus, we underscore how such platforms are viable architectural options for heterogeneous edge AI accelerators to complement ultra-low-power microcontrollers.